\documentclass[aps,prb,twocolumn]{revtex4}
\usepackage{amsmath}
\usepackage{graphicx}
\begin{document}

\title{Spin-Transfer Torques in Helimagnets}

\author{Kjetil M. D. Hals and Arne Brataas}
\affiliation{ Department of Physics, Norwegian University of Science and Technology, NO-7491, Trondheim, Norway }

\begin{abstract}
We theoretically investigate current-induced magnetization dynamics in chiral-lattice helimagnets. 
Spin-orbit coupling in non-centrosymmetric crystals induces a reactive spin-transfer torque that has not been previously considered. 
We demonstrate how the torque is governed by the crystal symmetry and acts as an effective magnetic field along the current direction in cubic B20-type crystals.
The effects of the new torque are computed for current-induced dynamics of spin-spirals and the Doppler shift of spin waves.   
In current-induced spin-spiral motion, the new torque tilts the spiral structure. 
The spin waves of the spiral structure are initially displaced by the new torque, while the dispersion relation is unaffected.     
\end{abstract}
\maketitle

\newcommand{\eq}  {  \! = \!  }
\newcommand{\keq} {\!\! = \!\!}
\newcommand{\kadd}{  \! + \!  }
\newcommand{\ksim}{\! \sim \!}

\section{Introduction}
Current-induced magnetization dynamics continue to be a very active research area due to potential applications in future electronic devices. In metallic ferromagnets, the magnetization can be manipulated via the spin-transfer torque (STT), which arises due to a misalignment between the spin polarization of the current and the local magnetization direction.~\cite{Ralph:jmmm08,Brataas:nature2012} Slonczewski and Berger were the first to predict the existence of the STT effect,~\cite{Berger,Slon} which was later demonstrated in several experiments.~\cite{Ralph:jmmm08,Brataas:nature2012} The anticipated application potential of the STT effect lies in the development of electro-magnetic devices that utilize a current-induced torque instead of external magnetic fields to manipulate the magnetization.~\cite{Brataas:nature2012}

The magnetization dynamics of an itinerant ferromagnet is described by the Landau-Lifshitz-Gilbert (LLG) equation extended to include the current-induced torques:~\cite{Brataas:nature2012,footnote1}
\begin{equation}
 \mathbf{\dot{m}} = - \gamma \mathbf{m}\times \mathbf{H}_{\rm eff} + \alpha\mathbf{m}\times \mathbf{\dot{m}} + \boldsymbol{\tau} . \label{Eq:LLG}
\end{equation}
Here, $\mathbf{m}= \mathbf{M}/ M_s$ ($M_s= |\mathbf{M}|$) is the unit direction vector of the magnetization $\mathbf{M}$, $\mathbf{H}_{\rm eff}= - \delta F / \delta \mathbf{M}$ is the effective field found by varying the free energy $F[\mathbf{M}]$ with respect to the magnetization,
$\alpha$ is the Gilbert damping coefficient, $\gamma$ is (minus) the gyromagnetic ratio, and $\boldsymbol{\tau}$ describes the current-induced torques.
In the absence of intrinsic spin-orbit coupling (SOC), the torque becomes $\boldsymbol{\tau}=\boldsymbol{\tau}_{\rm ex}$:~\cite{Brataas:nature2012}
\begin{equation}
\boldsymbol{\tau}_{\rm ex}= -(1 - \beta\mathbf{m}\times)(\mathbf{v}_s\cdot\boldsymbol{\nabla} )\mathbf{m} . \label{Eq:torque_metallic}
\end{equation}
In Eq.~\eqref{Eq:torque_metallic}, the first term is the adiabatic torque, while the second term (parameterized by  $\beta$) is the non-adiabatic torque. The vector $\mathbf{v}_s$ is proportional to the current density, $\boldsymbol{\mathcal{J}}$,  and its polarization, $P$: $\mathbf{v}_s = -\hbar P\boldsymbol{\mathcal{J}}/ 2e s_0$. 
Here, $s_0$ is the total equilibrium spin density along $-\mathbf{m}$, and $e$ is the electron charge.
 The torque in Eq.~\eqref{Eq:torque_metallic} treats the ferromagnet within the exchange approximation, which assumes that the exchange forces only depend on the relative orientation of the spins. This assumption is believed to be valid in metallic ferromagnets, including disordered systems in which impurities couple to the spin degree of freedom through random magnetic moments or spin-orbit coupling. In this case, impurity averaging restores the spin-rotational symmetry of the system.  
 Recently, in systems with a broken spatial inversion symmetry, the intrinsic SOC in combination with an external electric field have been observed to induce an additional torque,  
 such that $\boldsymbol{\tau}= \boldsymbol{\tau}_{\rm ex} + \boldsymbol{\tau}_{\rm so} $.~\cite{Manchon:prb08,Chernyshov:nature09,Garate:prb09,Hals:epl10,Miron:nature10} 
In general, the SOC-induced torque, $\boldsymbol{\tau}_{\rm so}$, can be written as~\cite{Manchon:prb08,Garate:prb09,Hals:epl10} 
\begin{equation}
\boldsymbol{\tau}_{\rm so}= -\gamma\mathbf{m}\times \mathbf{H}_{\rm so},\label{Eq:torque_SOC}
\end{equation}
where the SOC field, $\mathbf{H}_{\rm so}$, is proportional to the electric field and its orientation is determined by the symmetry of the underlying crystal lattice and the direction of the external electric field.  
Therefore, in contrast to $\boldsymbol{\tau}_{\rm ex}$, which vanishes in a homogeneous ferromagnet,  $\boldsymbol{\tau}_{\rm so}$ is finite even in this case. Several experiments have demonstrated that the SOC torque plays an important role in magnetization dynamics.~\cite{Chernyshov:nature09,Miron:nature10}  
The underlying physics of the torque is that the SOC effectively acts as a magnetic field on the spins of the itinerant quasi-particles when an electric field is applied to the system. The effective magnetic field induces an out-of-equilibrium spin density that yields a torque on the magnetization.~\cite{Manchon:prb08,Chernyshov:nature09}

In chiral magnets, the exchange interaction also contains an anisotropic term known as the Dzyaloshinskii-Moriya (DM) interaction.~\cite{Dzyaloshinsky:jpcs58,Moriya:pr60}
The DM interaction arises due to the characteristic crystalline asymmetry of the chiral magnet in combination with the SOC, and in cubic B20-type crystals, it leads to the formation of a spin spiral in the magnetic ground state. 
We refer to these systems as helimagnets.
Helimagnets have recently attracted substantial interest because topological nontrivial spin structures, known as skyrmions, have been observed in such systems under the application of weak external magnetic fields.~\cite{Rossler:nature06,Muhlbauer:science09,Yu:nature10,Yu:natmat11,Tonomura:nanolett12,Seki:science,Kanazawa:science} 
Current-induced responses of the formed skyrmion lattice to current densities that are over five orders of magnitude smaller than those typically observed in conventional ferromagnetic metals have recently been observed experimentally.~\cite{Jonietz:science10,Yu:natcom12} 
To understand this striking feature of helimagnets, numerical simulations and a collective coordinate description have been applied to study the current-induced dynamics of spin-spirals and skyrmion lattices.~\cite{Goto:arxiv08,Zang:prl11,Iwasaki:arxiv12} 
However, an important aspect of helimagnets is the absence of spatial inversion symmetry, which implies that the magnetization experiences a SOC-induced torque given by Eq.~\eqref{Eq:torque_SOC}. None of the recent theoretical works have included or derived the explicit form of the term in Eq.~\eqref{Eq:torque_SOC}. Its particular form and importance for the current-induced dynamics of helimagnets therefore remain unknown.

In the present paper, we derive the form of the torque in Eq.~\eqref{Eq:torque_SOC} for cubic non-centrosymmetric (B20-type) compounds. An important example of such a system is the chiral itinerant-electron magnet MnSi, which was the first system in which a two-dimensional lattice of skyrmions was observed. The effects of the SOC torque are studied for two different cases: current-induced spin-spiral dynamics and the Doppler shift of magnons that propagates along the spiral structure. We observe that for current-induced spin-spiral motion, the new torque yields an enhanced tilting of the spiral structure, while the torque does not affect the Doppler shift of spin waves except to induce an initial translation of the spiral structure. 
We also briefly discuss the effect of the SOC torque on the skyrmion lattice dynamics.

This paper is organized in the following manner. Sec.~\ref{Sec:SOC_torque} provides a derivation of the SOC torque in Eq.~\eqref{Eq:torque_SOC} for cubic B20-type crystals. Sec.~\ref{Sec:Results} discusses the effects of the SOC torque on current-induced spin-spiral motion and the Doppler shift of spin waves that propagates along the spin-spiral. We conclude and summarize our results in Sec.~\ref{Sec:Summary}.

\section{Derivation of the SOC torque} \label{Sec:SOC_torque}
In deriving the explicit form of the torque in Eq.~\eqref{Eq:torque_SOC}, we are guided by the Onsager reciprocity relations and Neumann's principle. Consider a system 
described by the parameters $\{ q_i|i=1,\ldots,N\}$ for which the rate of change $\dot{q}_i$ is induced by the thermodynamic forces $f_i\equiv-\partial F/ \partial q_i$, where $F(q_1, ..., q_N)$ is the free energy.
Onsager's theorem states that the response coefficients in the equations $\dot{q}_i =\sum_{j=1}^{N} L_{ij}f_j$ are related by $L_{ij}(\mathbf{H},\mathbf{M})= \epsilon_i\epsilon_jL_{ji}(-\mathbf{H},-\mathbf{M})$, where $\epsilon_i=1$ ($\epsilon_i=-1$) if $q_i$ is even (odd) under time reversal.~\cite{Birss:book} $\mathbf{m}$ and $\mathbf{H}$ represent any possible equilibrium magnetic order and an external magnetic field, respectively. 
In the present paper, the responses of the itinerant ferromagnet are described by the time derivative of the unit vector along the magnetization direction, $\dot{\mathbf{m}}$,  and the charge current density, $\boldsymbol{\mathcal{J}}$. The associated 
thermodynamic forces are the effective field scaled with the magnetization, $\mathbf{f}_{\mathbf{m}}= M_s \mathbf{H}_{\rm eff}$, and the electric field, $\mathbf{f}_{\mathbf{q}}= \mathbf{E}$, respectively,
and the equations describing the dynamics in the linear response regime are determined by:
\begin{equation}
\begin{pmatrix}
\mathbf{\dot{m}} \\
\boldsymbol{\mathcal{J}}
\end{pmatrix} =
\begin{pmatrix}
\mathbf{L}_{\mathbf{m}\mathbf{m}} & \mathbf{L}_{\mathbf{m}\mathbf{q}}  \\
\mathbf{L}_{\mathbf{q}\mathbf{m}} & \mathbf{L}_{\mathbf{q}\mathbf{q}}
\end{pmatrix}
\begin{pmatrix}
\mathbf{f}_{\mathbf{m}} \\
\mathbf{f}_{\mathbf{q}}
\end{pmatrix} . \label{Eq:Onsager}
\end{equation}
The Onsager reciprocity relations imply that $L_{m_iq_j} (\mathbf{m}) = -L_{q_jm_i} (-\mathbf{m})$.
In addition to the symmetry requirements imposed by the reciprocity relations, the symmetry of the underlying lattice structure also decreases the number of independent tensor components.
This fact is expressed by Neumann's principle, which states that a tensor representing any physical property should be invariant with respect to every symmetry operation of the crystal's point group.~\cite{Birss:book}     

According to Eq.~\eqref{Eq:Onsager}, the effect reciprocal to the adiabatic and non-adiabatic torque in Eq.~\eqref{Eq:torque_metallic} is a charge current density induced by a time-dependent magnetic texture.
To the lowest order in the texture gradients and the precession frequency, the induced charge current density in the exchange approximation is~\cite{Tserkovnyak:prb08}
\begin{equation}
\mathcal{J}_i^{{\rm ex}} = \frac{\hbar}{2e}\sigma P \left( \mathbf{m}\times \frac{\partial \mathbf{m}}{\partial r_i} - \beta\frac{\partial \mathbf{m}}{\partial r_i}  \right)\cdot \mathbf{\dot{m}} .
\end{equation}
Here, $e$ is the electron charge, $P$ is the spin polarization of the current,  $\sigma$ is the conductivity, and $r_i$ is component $i$ of the spatial vector. 
Because the exchange approximation neglects any coupling (via intrinsic SOC) of the spins to the crystal structure, the above expression is fully spin-rotational symmetric and a textured
magnetization, i.e., $\partial \mathbf{m} / \partial r_i \neq 0$, is required to have a coupling between the momentum of the itinerant quasiparticles and the magnetization.  
If the effects of intrinsic SOC are considered, additional terms are allowed by symmetry in the phenomenological expansion for the pumped current. 
In particular, for inversion symmetry-breaking SOC, a homogenous magnetization pumps a charge current. 
To the lowest order in SOC and precession frequency, the expression for the pumped current then becomes  
\begin{equation}
\mathcal{J}_i^{{\rm pump}} = \eta_{ij} \dot{m}_j  +  \mathcal{J}_i^{{\rm ex}} . \label{Eq:j_pump}
\end{equation}
The second-rank tensor $\eta_{ij}$ is an axial tensor because the current is a polar vector while the magnetization is an axial vector. 
$\eta_{ij}$ is linear in the SOC coupling and vanishes in systems with spatial inversion symmetry. 
According to Neumann's principle, the particular form of $\eta_{ij}$ is governed by
the crystal structure and is determined by the following set of equations produced by the generating matrices $[R_{ij}]$ of the crystal's point group:~\cite{Birss:book}
\begin{equation}
\eta_{ij}= |R| R_{in} R_{jm} \eta_{nm} . \label{Eq:Neumann}
\end{equation}
Here, $|R|$ is the determinant of the matrix $[R_{ij}]$. 

Let us now consider a cubic B20-type crystal. Its crystal structure belongs to the non-centrosymmetric space group $P2_13$, which has the cubic point group $T$. Common examples of cubic B20-type chiral magnets are MnSi, FeGe, and (Fe,Co)Si.
From the symmetry relations in Eq.~\eqref{Eq:Neumann}, one then finds
that the tensor $\eta_{ij}$ is proportional to the unit matrix:~\cite{Birss:book}
\begin{equation}
 \eta_{ij}= \eta \delta_{ij}, 
 \end{equation}
 where $\delta_{ij}$ is the Kronecker delta. 
 The tensor is isotropic because the high symmetry of the cubic crystal reduces the number of independent tensor coefficients to the single parameter $\eta$.
 Substituting this tensor into Eq.~\eqref{Eq:j_pump} and expressing the time derivative of the magnetization in terms of  
 the effective field by applying the first term on the right-hand side of Eq.~\eqref{Eq:LLG}, one obtains the response matrix:~\cite{comment1}  
\begin{equation}
L_{q_i m_j} =
-\frac{\gamma\eta}{ M_s}\epsilon_{ikj}m_k  + L_{q_i m_j}^{\rm ex}.
\end{equation} 
Here, $L_{q_i m_j}^{\rm ex}$ are the response coefficients describing the process reciprocal to the STT in Eq.~\eqref{Eq:torque_metallic}, which have been previously derived in Ref.~\onlinecite{Tserkovnyak:prb08}. The term
proportional to $\eta$ describes the process reciprocal to the SOC-induced torque in Eq.~\eqref{Eq:torque_SOC}. Using the Onsager reciprocity relations, we find that  the  SOC field takes the following form:
\begin{equation}
\mathbf{H}_{\rm so} = \eta_{\rm so} \mathbf{v}_s, \label{Eq:H_soc}
\end{equation}
where $\eta_{\rm so} = (2\eta e s_0 )/(\hbar \sigma P M_s )$.
Thus, the torque induced by the SOC in non-centrosymmetric  cubic magnets acts as an effective magnetic field along the current direction.
Note that the torque is reactive because it does not break the time reversal symmetry of the LLG equation.

\section{Results and discussion} \label{Sec:Results}
 In the following, we investigate the effects of the SOC torque on current-driven spin-spiral motion and the Doppler shift of spin waves.
 Additionally, a brief discussion of how we expect the torque to affect the skyrmion crystal dynamics is presented.    
 
\subsection{Spin-Spiral Motion}
To the lowest order in the magnetic texture gradients, the free energy density of a ferromagnet with broken spatial inversion symmetry can be written phenomenologically as:~\cite{LL:book}
\begin{equation}
\mathcal{F}(\mathbf{m}) = \frac{J_{ij}}{2} \frac{\partial \mathbf{m}}{\partial r_i}\cdot \frac{\partial \mathbf{m}}{\partial r_j} + D_{ijk} m_i \frac{\partial m_j}{\partial r_k} .  \label{Eq:FreeE_0}
\end {equation}
Here, $J_{ij}$ is the spin stiffness describing the exchange interaction between neighboring magnetic moments, and the term proportional to $D_{ijk}$ is the DM interaction. In Eq.~\eqref{Eq:FreeE_0} (and in what follows),  summation 
over repeated indices is implied. The explicit form of the tensors $J_{ij}$ and $D_{ijk}$ is determined by the crystal symmetry. 

In cubic B20-type ferromagnets, the free energy density becomes  
\begin{equation}
\mathcal{F}(\mathbf{m}) = \frac{J}{2} \frac{\partial \mathbf{m}}{\partial r_i}\cdot \frac{\partial \mathbf{m}}{\partial r_i} + D \mathbf{m}\cdot\left( \boldsymbol{\nabla}\times\mathbf{m} \right). \label{Eq:FreeE}
\end {equation}
The free energy of the system, $F\left[ \mathbf{m}\right] = \int d\mathbf{r}\mathcal{F}$, is minimized by a helical magnetic order, where the wave vector of the spiral structure is determined by the ratio between the DM parameter and
the spin stiffness: $k = D/J$. For a $\mathbf{k}$-vector that points along the $z$ axis, the magnetic order of the ground state is
\begin{equation}
\mathbf{m}_0 (z) = \cos (kz)\hat{\mathbf{x}}  + \sin (kz) \hat{\mathbf{y}},
\end{equation}         
where $\hat{\mathbf{x}}$ and $\hat{\mathbf{y}}$ are the unit direction vectors along the $x$ and $y$ axis, respectively. 

The action functional, $S[\mathbf{m}]$, and the dissipation functional, $R[\mathbf{\dot{m}}]$, of the system are written as~\cite{Auerbach:book,Gilbert:2004}
\begin{eqnarray}
S \left[ \mathbf{m} \right] &=& \int dtd\mathbf{r}~A_i \left(  \dot{m}_i  + \mathbf{v}_s\cdot\boldsymbol{\nabla} m_i \right)  + \nonumber \\ 
&& \frac{\gamma}{M_s}   \mathcal{F}(\mathbf{m}) - \gamma \mathbf{m}\cdot\mathbf{H}_{\rm so}   ,  \label{Eq:S}  \\
R\left[ \mathbf{\dot{m}} \right] &=&  \int dtd\mathbf{r}~ \frac{\alpha}{2} \left(    \mathbf{\dot{m}}  + \frac{\beta}{\alpha}\mathbf{v}_s\cdot\boldsymbol{\nabla}\mathbf{m}   \right)^2 . \label{Eq:R}
\end{eqnarray}
Here, $\mathbf{A} (\mathbf{m})$ is the Berry phase vector potential of a magnetic monopole, which satisfies $ \epsilon_{ijk}\partial A_k / \partial m_j = m_i$ ($\epsilon_{ijk}$ is the Levi-Civita tensor).
The LLG equation in Eq.~\eqref{Eq:LLG},  with 
$\boldsymbol{\tau}= \boldsymbol{\tau}_{\rm ex} + \boldsymbol{\tau}_{\rm so}$, is determined by
\begin{equation}
\frac{\delta S }{\delta \mathbf{m}} = -\frac{\delta R }{\delta \mathbf{\dot{m}} } . \label{Eq:LLG_2}
\end{equation} 

A previous study on spin-spiral motion demonstrated that the response of the structure to an applied current (along $z$) can be described by the tilting angle, $\xi$, and drift velocity, $\dot{\zeta}$, of the spiral structure.~\cite{Goto:arxiv08}    
To find an approximate solution of Eq.~\eqref{Eq:LLG_2}, we therefore employ the following variational ansatz:
\begin{equation}
\mathbf{m} (z,t) = \cos(\xi (t))\mathbf{m}_0 (z - \zeta (t) ) + \sin (\xi (t))\hat{\mathbf{z}} . \label{Eq:ansatz}
\end{equation} 
Substitution of this ansatz into Eq.~\eqref{Eq:S} and \eqref{Eq:R} and integration over the spatial coordinates yield an effective action and dissipation functional for the variational parameters $\xi (t)$ and $\zeta (t)$:
\begin{eqnarray}
S [\zeta ,\xi] &=& \int dt   \left(  \dot{\zeta} - v_s \right) k  \sin{\xi}  +  \label{Eq:S_eff} \\
& & \frac{\gamma}{M_s} \left(  \frac{J}{2} k^2\cos^2\xi - Dk \cos^2\xi  \right) - \gamma H_{\rm so}\sin\xi   ,   \nonumber  \\
R [\dot{\zeta},\dot{\xi}] &=& \int dt \frac{\alpha}{2} \left[ \dot{\xi}^2 +  \left( \frac{\beta}{\alpha}v_s k - k\dot{\zeta}  \right)^2  \right] .
\end{eqnarray}
The equations of motion for the variational parameters are 
\begin{eqnarray}
\frac{\delta S [\zeta ,\xi] }{ \delta \zeta} = -\frac{\delta R [\dot{\zeta},\dot{\xi}] }{ \delta \dot{\zeta}} & ,& \frac{ \delta S [\zeta,\xi] }{ \delta \xi} = -\frac{\delta R [\dot{\zeta},\dot{\xi}] }{ \delta \dot{\xi} }  .
\end{eqnarray}
We are interested in the steady-state regime in which  $\xi$ approaches a constant value. In this regime,  the drift velocity and the tilting angle are
\begin{eqnarray}
\dot{\zeta}  &=& \frac{\beta}{\alpha} v_s ,  \\
\sin (\xi) &=& \frac{M_s}{\gamma} \frac{1}{Jk - 2D}\left(  \left( \frac{\beta}{\alpha} - 1 \right) v_s - \frac{\gamma}{k } H_{\rm so} \right) .
\end{eqnarray} 
The expression for the drift velocity $\dot{\zeta}$ agrees with the expression derived in Ref.~\onlinecite{Goto:arxiv08}.   
The SOC torque does not affect the drift velocity because  the SOC torque effectively acts similarly to the adiabatic torque, as can be observed from  
the expression for the action $S [\zeta ,\xi] $ in Eq.~\eqref{Eq:S_eff}.  The adiabatic and SOC torques  
initiate a spiral motion when a current is applied. However, the motion is damped due to the intrinsic pinning effect caused by Gilbert damping in combination with the DM interaction.
Thus, similar to what is observed for domain walls in conventional ferromagnets, a non-adiabatic torque is required to obtain a steady-state spiral motion. 
An observable effect of the SOC torque is the modification of the tilting angle observed in Ref.~\onlinecite{Goto:arxiv08} by an amount of $-M_s H_{\rm so}  / (Jk^2-2Dk) $.

\subsection{Doppler Shift of Spin Waves}
In ferromagnets with a homogeneous magnetization,  a Doppler shift in the spin wave dispersion relation under the application of a current has been observed.~\cite{Vlaminck:science08}
The frequency, $\omega$, of the spin wave is shifted by $\mathbf{v}_s\cdot \mathbf{q}$, where $\mathbf{q}$ is the wave vector of the magnon:
$\omega = (\gamma J / M_s)\mathbf{q}^2 + \mathbf{v}_s\cdot \mathbf{q}$. 

Theoretical works on Goldstone modes in helimagnets with a spin spiral predict that these modes are much more complicated than those in  
homogeneous ferromagnets.~\cite{Belitz:prb06}  We refer to these Goldstone modes as helimagnons. 
The dispersion relation of the helimagnons is highly anisotropic, with a linear wave-vector dependency parallel to the spin-spiral direction and a quadratic dependency 
in the transverse direction (in the long wave length limit). That is, the soft modes behave like antiferromagnetic magnons along the spiral, while ferromagnetic behavior is
observed for modes propagating in the transverse plane. Thus far, no works have studied the effect of an applied current on the dispersion relation of helimagnons.   

To derive an effective action for the Goldstone modes, we describe the local fluctuations by $\xi$ and $\zeta$ in Eq.~\eqref{Eq:ansatz} by allowing the parameters to be both position- and time-dependent:
$\xi= \xi (\mathbf{r},t)$ and $\zeta= \zeta (\mathbf{r},t)$. 
A similar parameterization was performed in Refs.~\onlinecite{Belitz:prb06,Petrova:prb11} in the analysis of helimagnons.
The parameter $\zeta$ describes a local twist (around the $z$ axis) of the spiral structure, while $\xi$ describes 
a local tilting along the $z$ axis. In the analysis of the Doppler shift, we neglect dissipation and disregard the dissipation function.  
Ref.~\onlinecite{Petrova:prb11} demonstrated that simple closed-form solutions for the 
variational parameters can only be obtained for modes propagating along the spin-spiral direction. 
For simplicity, we therefore restrict our study to Goldstone modes that propagate along the $z$ axis. 
Expanding Eq.~\eqref{Eq:S} to second order in $\xi (z,t)$ and $\zeta (z,t)$, we obtain the effective action (the current is applied along the $z$ axis):
\begin{eqnarray}
S\left[ \xi, \zeta \right] &=& \int dtdz~  k\xi \left( \dot{\zeta}  + v_s  \frac{\partial \zeta}{\partial z} - v_s \right) +   \\
  & & \frac{\gamma J}{2M_s} \left[  \left( \frac{\partial \xi }{\partial z} \right)^2 + k^2 \left( \frac{\partial \zeta }{\partial z} \right)^2  +  k^2\xi^2  \right] - \gamma H_{\rm so}\xi    .  \nonumber 
\end{eqnarray}
The equations of motion are obtained by varying the action with respect to $\xi$ and $\zeta$, i.e., $\delta S/ \delta \zeta = \delta S/ \delta \xi = 0$, which
results in two coupled equations for the variational parameters:
\begin{eqnarray}
\dot{\zeta}(z,t)  + v_s \frac{\partial \zeta (z,t)}{\partial z} &=& -\frac{\gamma J}{k M_s}\left(  k^2  - \frac{\partial^2}{\partial z^2} \right) \xi (z,t) + \nonumber  \\
&& v_s + \frac{\gamma }{k } H_{\rm so} , \label{Eq:zeta}  \\
\dot{\xi}(z,t)  + v_s \frac{\partial \xi (z,t)}{\partial z}  &=& -\frac{\gamma J k}{M_s} \frac{\partial^2 \zeta (z,t)  }{\partial z^2}   .\label{Eq:xi}
\end{eqnarray}
Let us first consider the homogenous part of the equations and neglect the two last terms on the right-hand side in Eq.~\eqref{Eq:zeta}.
Substitution of a plane wave ansatz of the form $\left[ \zeta_0 \  \xi_0 \right]^T\exp (i (qz - \omega t) )$ into the equations yields the following
dispersion relation:
\begin{equation}
\omega = \frac{\gamma J}{M_s} q \sqrt{k^2 + q^2} + v_s q.  
\end{equation}    
We see that the STT results in a Doppler shift similar to what is observed for spin waves in conventional ferromagnets.  
In the long wave length limit, $q\rightarrow 0$, a linear dispersion relation is obtained: $\omega = (\gamma J/ M_s)k q + v_s q $.  
The SOC-induced torque only appears as a source term in the nonhomogeneous  equations.
The particular solution (PS) of the nonhomogeneous equations in Eq. \eqref{Eq:zeta} and \eqref{Eq:xi} is 
\begin{equation}
\begin{pmatrix}
\zeta(z,t) \\
\xi(z,t)
\end{pmatrix}_{\rm PS}
= 
\begin{pmatrix}
(v_s + (\gamma/ k)H_{\rm so})t  \\
0
\end{pmatrix} .
\end{equation}
This solution describes a displacement of the spiral structure induced by the adiabatic and SOC torques. However, this current-driven spin-spiral motion is damped when dissipation 
is considered due to the intrinsic pinning effect. Thus, the SOC torque (together with the adiabatic torque) only causes an initial translation of the spin-spiral.

\subsection{Skyrmion Crystal Dynamics} 
In helimagnetic thin-film systems, skyrmions have been observed under the application of a weak external magnetic field, $\mathbf{B}$,  perpendicular to the thin film.
Each skyrmion has a vortex-like magnetic configuration, where the magnetic moment at the core of the vortex is anti-parallel to the applied field while 
the peripheral magnetic moments are parallel. From the peripheral moments to the core, the magnetic moments swirl up in a counter-clockwise or clockwise manner.   
The formed skyrmions arrange themselves in a crystalline structure, a two-dimensional skyrmion crystal.

Recent experiments have revealed curren-driven skyrmion crystal motion at ultra low current densities.\cite{Jonietz:science10} 
The motion of a skyrmion lattice is only weakly affected by pinning, which is in stark contrast to observations for
current-induced domain wall dynamics in conventional ferromagnets. A theoretical work has indicated that the pinning-free motion arises because
the skyrmion lattice rotates and deforms to avoid the impurities.~\cite{Iwasaki:arxiv12} However, all analyses of current-driven skyrmion crystal motion have disregarded
the SOC torque.

 Sec.~\ref{Sec:SOC_torque} showed that the SOC torque acts as an effective field along the current direction.  For a current applied along any direction
in the thin film, the expected consequence of the SOC torque is therefore that this torque leads to a small correction to the external magnetic field that stabilizes the two-dimensional skyrmion lattice
such that the total field becomes $\mathbf{H}_{\rm T}= \mathbf{B} + \mathbf{H}_{\rm so}$. The expected response of the skyrmion crystal to this perturbation is a rotation of the 2D lattice structure
that aligns the core magnetic moments anti-parallel to $\mathbf{H}_{\rm T}$.  To confirm our predictions, a more thorough numerical simulation of the magnetic system is required, which is beyond
the scope of the present work.

\section{Summary} \label{Sec:Summary}
In this paper, we performed a theoretical study of current-induced torques in cubic non-centrosymmetric helimagnets.
We demonstrated that due to the broken spatial inversion symmetry, the SOC induces a reactive 
magnetization torque that has not been considered in previous studies of current-induced dynamics
in helimagnets. The specific form of the SOC torque is determined by the symmetry of the underlying crystal lattice and  
acts as an effective magnetic field along the current direction in B20-type chiral magnets. 

The consequences of the SOC torque are studied for two different cases: 
current-induced spin-spiral motion and the Doppler shift of helimagnons.  
During the current-driven spin-spiral motion, the SOC torque yields an enhanced tilting of the spin-spiral structure, while the 
velocity is not affected. The dispersion relation of a helimagnon that propagates along the axis of the spin-spiral is not affected by the SOC torque 
except to induce an initial translation of the spiral structure.

\section{Acknowledgments}
This work was supported by EU-ICT-7 contract No. 257159 "MACALO".


\begin{thebibliography}{99}
\bibitem{comment1}We have disregarded the magnetization damping in the derivation the Onsager coefficients. If damping is included, the same results are obtained, but a transformation between the
Landau-Lifshitz equation and the Landau-Lifshitz-Gilbert equation is required at two intermediate steps in the derivation. See Refs.~\onlinecite{Tserkovnyak:prb08,Hals:epl10} for further details.
\end{thebibliography}
\end{document}